\documentclass[twoside]{sfm}

\input{sf.def}

\begin{document}

\pagebreak

\thispagestyle{titlehead}

\setcounter{section}{0}
\setcounter{figure}{0}
\setcounter{table}{0}

\markboth{Author1 et al.}{Magnetic fields along the PMS}

\titl{Magnetic fields along the pre-main-sequence phase}{Alecian E.$^{1,2}$}
{$^1$UJF-Grenoble 1 CNRS-INSU, Institut de Plan\'etologie et d'Astrophysique de Grenoble (IPAG) UMR 5274, Grenoble, F-38041, France, email: {\tt evelyne.alecian@obs.ujf-grenoble.fr} \\
 $^2$LESIA, Observatoire de Paris, CNRS, UPMC, Universit\'e Paris-Diderot, 5 place Jules Janssen, 92195 Meudon C\'edex, France
 }

\abstre{
In this contributoin I review some of the major work that have been undertaken in the last decade to study the magnetic fields in intermediate-mass pre-main sequence stars, with the goal of understanding the origin of the magnetic fields detected in the chemically peculiar Ap/Bp stars, and testing the fossil field theory. 
}

\baselineskip 12pt

\section{Introduction}

\subsection{Magnetic fields in the low-mass stars}

Magnetic fields in the cool low-mass stars are ubiquitous, highly variable, are often of complex configuration and sometimes show cycles, when they can be detected. These characteristics are very similar to the solar magnetic field, and by extension are assumed to be formed as in the sun, by a {\it solar-type dynamo}. Indeed, the sub-surface convective layers of these cool stars are convective, and the large plasma motions occurring in the convective cells, by interacting with the stellar differential rotation, can generate the complex magnetic fields observed at the surface. As a consequence a clear correlation between the rotation period of the star and the strength of the magnetic activity is observed. Besides, in the case of the main sequence low-mass stars and partially convective PMS low-mass stars, we observe the fields to be concentrated in small area of the surface of the star and to be associated with cool spots, such as the solar spots easily observable at the surface of the Sun, but also on other stars by variable photometry. These fields can also have a large impact on the evolution of the stars during the various phases of formation and evolution, by interacting with their close environment via e.g. jet launching, magnetised winds, magnetic disc locking, or magnetospheric accretion (see the recent review on stellar magnetic properties in \cite{donati09}).

\subsection{Magnetic fields in the Ap/Bp stars}

In the higher-mass stars ($M>1.5$M$_{\odot}$), on the main sequence, magnetic fields are also observed in some A and B stars, but with very different characteristics. The fields are strong (from 300 G to about 30 kG), rare (less than 10 \% of the A and B stars), they are organised on large scale, i.e. they are mainly dipolar with sometimes a quadrupolar and/or an octupolar component, they are stable over many years, and even decades for the stars that have been observed that long, they are not correlated with any stellar properties, including the rotation period or the age, they are all found in chemically peculiar Ap/Bp stars, and many are found in slow rotators \cite{donati09}. All these characteristics being very different from the cool stars magnetic properties, it implies a different origin. In the last couple of decades, two main theories have been proposed : the core-dynamo theory and the fossil field theory. A third one has been recently proposed: the merger theory, but I refer the reader to G. Mathys' contribution of this volume.

\subsection{The core dynamo theory}

The main sequence A and B stars interiors are constituted with a convective core surrounded with a large radiative envelope. It has been proposed that, as in the convective envelope of the sun, a magnetic field could be generated in the core and then diffuses over the radiative envelope to produce the fields observed at the surface of some A and B stars \cite{schussler78}. However this hypothesis is highly unlikely for many reasons \cite{moss01}: 1) solar-type dynamo fields have variable complex configuration, and it is not clear how complex variable fields generated in the core can produce stable large-scale organised fields at the surface ; 2) the field of the Ap/Bp stars are very strong and again it is not clear how dynamo field generated in the small core and then diluted over the large stellar surface can produce such strong fields ; 3) in Ap/Bp stars we observe very strong fields ($B_{\rm P} \sim10$ kG or more) in slow rotators (rotation periods larger than few days, e.g. \cite{bailey12}), while the solar-dynamo theory, due to the role that plays differential rotation, predicts the faintest fields in the slowest rotators (e.g. \cite{noyes84}); 4) the diffusion time for the field to reach the surface is equal to a significant fraction of the main-sequence lifetime, and is therefore inconsistent with the detection of fields in stars that have just formed (e.g. \cite{alecian08a}, \cite{alecian08b}) ; 5) fossil fields have been detected in totally radiative pre-main sequence stars (e.g. \cite{catala07}) ; 6) similar fields are observed in stars with very different properties (different mass, temperature, or age), and fields of different configuration or strength are observed in similar stars (e.g. \cite{petit13}); 7) fields are only observed in a small fraction of the A and B stars \cite{power08}, while all of them have a convective core and should therefore be magnetic. To summarise, it appears that the core dynamo theory can explain none of the magnetic properties observed in the A and B stars. 

\subsection{The fossil field theory}


All magnetic properties observed in the high-mass stars favour the fossil field theory, i.e. a magnetic field that is not being continuously maintained against ohmic decay \cite{borra82}. This field is assumed to have been shaped during the star formation. The details on the physical origin are still not clear, but I will propose here two scenario. The first one assumes that the Galactic magnetic fluxes, threading the parental molecular clouds in which stars form, are swept up during the star formation to be concentrated into the protostars, to finally become the large-scale and slowly decaying magnetic fields observed at the surface of the MS stars.  The second scenario considers a magnetic field being generated during the fully convective phases of the proto-stellar evolution of intermediate-mass stars. These newly generated fields would relax into a large-scale configuration inside the radiative interiors, as soon as they take place during the pre-main sequence phase. 

\subsection{The PMS evolution}

\begin{figure}[!t]
\begin{center}
\includegraphics[width=11cm]{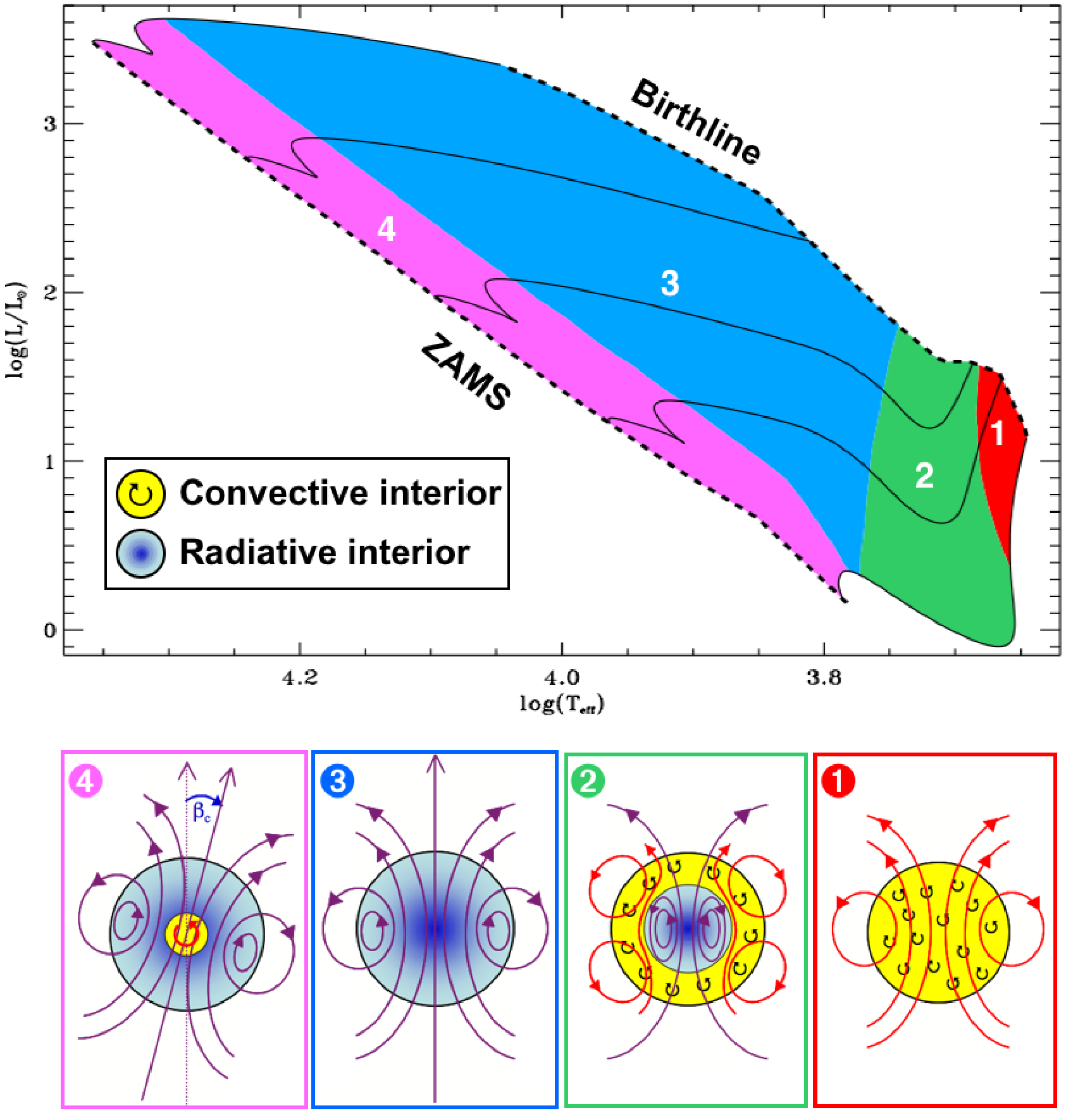}
\vspace{-5mm}
\caption[]{{\bf Top:} Pre-main sequence evolutionary tracks (full lines) for 1.2, 2, 3, 5, and 8 M$_{\odot}$, computed with CESAM (Morel 1997). The birthline is from Behrend \& Maeder (2001). The shaded area separate regions with different stellar internal structures: phase 1 (red), fully convective interior ; phase 2 (green) radiative core + convective envelope ; phase 3 (blue), fully radiative interior ; phase 4 (pink), convective core +  radiative envelope. {\bf Bottom:} Schematic view of the different natures of the magnetic fields for the four different phases. Yellow with circular arrows regions represent convective interior creating dynamo fields (represented with red field lines). Blue regions represent radiative interiors containing large-scale fossil fields (represented with purple field lines). }
\label{fig:hr}
\end{center}
\end{figure}

The pre-main sequence evolution is defined as the last period of stellar formation after the proto-stellar phase and before the main sequence phase. At the beginning of the PMS phase, the star is on the birthline, i.e. the locus in the HR diagram where the star is for the first time observable in the optical domain after getting rid of the majority of its proto-stellar envelope, once accretion stopped \cite{stahler83}. On the birthline, the star has not yet started nuclear reactions in its core. Then the star evolves along the PMS track in a quasi-static gravitational contraction, which is its main source of energy. Just before the end of the PMS phase, nuclear reaction are slowly starting, and contribute more and more to the luminosity of the star. The contraction is slowly coming to a stop until the star reaches the Zero-Age-Main-Sequence (ZAMS) (Fig. \ref{fig:hr}).

Palla \& Stahler \cite{palla90} have computed the birthline at intermediate-mass and find that a unique proto-stellar mass accretion rate of $10^{-5}$~M$_{\odot}.$yr$^{-1}$ gives rise to a theoretical birthline that reproduces well the upper envelope of the distribution of Herbig Ae/Be stars in the HR diagram. They also find that above 8~M$_{\odot}$ their birthline intersect the ZAMS, and conclude that no pre-main sequence stars more massive than 8~M$_{\odot}$ should be visible in the optical domain. However Alecian et al. \cite{alecian13b} argue that the recent discoveries of pre-main sequence stars with masses larger than 8~M$_{\odot}$ indicate that a unique mass accretion rate at all masses might not be fully correct, and used instead the birthline computed by Behrend \& Mader (\cite{behrend01}, BM01 hereafter), which reproduces well the upper envelope of the distribution of Herbig Ae/Be stars in the HR diagram, when including the most massive recently discovered ones. Behrend \& Mader used instead a growing accretion rate to compute the birthline. They assume that all stars, whatever their final mass, is evolving along the birthline from the lower right to the upper left during the proto-stellar phase, i.e. the phase during which the accretion is dominating. The star is therefore still embedded in its parental cloud accreting a large amount of material with an accretion rate that depends on the luminosity of the embedded star. Once the star has accreted most of the surrounding material, i.e. has depleted most of its close environment it leaves the birthline to follow the PMS tracks. In this simple view, all stars, whatever their mass, have been formed from a unique core of a certain mass, and what determines the final mass of the star is the mass of the original cloud from which the star forms.

During their PMS evolution the energy transport inside the star is changing. At lower-mass, stars are starting the PMS phase with a fully convective interior, and, while contracting, follow the PMS evolution along the Hayashi track, i.e. the almost vertical portion of the track. At one point the star has contracted enough so the energy is more efficiently transported by radiation, and a radiative core is forming and growing until reaching the surface. Once the nuclear reactions are triggered a convective core is forming, right before reaching the ZAMS (Fig. \ref{fig:hr}). At higher mass, the evolution is similar, but they pass the convective phases on the brithline and start their PMS evolution either partially radiative or totally radiative. I should point out here, that according to this scenario, all stars, whatever their final mass, go through a fully convective phase, before or at the beginning of the PMS phase. This fully convective phase should therefore be considered seriously in the fossil field scenario, as convective motion can have a high impact on the generation or the diffusion of magnetic fields. In the following I will review the project that have been undertaken the last decade, and that are still going on, in order to determine the magnetic properties of the PMS intermediate-mass stars during the four major phases of the PMS evolution (i.e. the fully convective phase, the core-radiative phase, the fully radiative phase, and the core-convective phase, respectively red, green, blue and pink in Fig. 1).

\section{Testing the fossil field theory}

\subsection{The Herbig Ae/Be survey}

We performed a high-resolution spectropolarimetric survey of 128 Herbig Ae/Be stars using the two most efficient instruments in the world: ESPaDOnS (Canada-France-Hawaii Telescope, Hawaii) and Narval (T\'elescope Bernad Lyot, France). We have selected the field Herbig Ae/Be stars in the catalogue of Th\'e et al. \cite{the94} and Vieira et al. \cite{vieira03}, while the Herbig Ae/Be stars members of young clusters have been selected in Park \& Sung (2002) for NGC 2244, Sung et al. (1997) for NGC 2264, and de Winter et al. (1997) for NGC 6611. Before this survey, the global magnetic properties of the Herbig Ae/Be stars were unknown. Some attempts to detect magnetic fields in few stars have been performed without much success (e.g. Catala et al. 1993, 1999), except in HD 104237 (Donati et al. 1997). Our survey allowed us to detect magnetic fields in 8 stars (e.g. Wade et al. 2005, Alecian et al. 2008b) bringing a magnetic incidence of about 6\%, similar to the incidence on the main sequence. Following this survey, we performed spectropolarimetric monitoring of many of the detected stars in order to characterise their magnetic fields, and we find that all of them have similar magnetic properties than on the main sequence: mainly dipolar, strong (300 G to 4 kG), and stable over many years (e.g. Alecian et al. 2008a, 2009). Thanks to this survey we have therefore established a fossil link between the PMS and the MS, and we conclude that the magnetic properties of the A/B stars must have been shaped before the Herbig Ae/Be phase of the stellar evolution (Alecian et al. 2013b).

\subsection{The stability of fossil fields}

The fossil field theory proposes that magnetic fields reside inside the stars without being continuously renewed. However, does an equilibrium state exist for a star hosting such a fossil field ? And can this equilibrium configuration be stable on an Alfv\'en timescale ? Prendergast (1956) has shown that a magnetic field can be stable inside a self-graviting axisymetric fluid only if toroidal and poloidal fields exist together. Numerical and analytical work have indeed shown that mixed fields (with both toroidal and poloidal components being non null) are stable inside radiative stars (Braithwaite \& Nordlund 2006, Duez et al. 2010). Furthermore they have also shown that random initial fields evolve towards a stable mixed configuration inside radiative stars (Braithwaite \& Spruit 2004, Duez \& Mathis 2010). Those theoretical work have therefore demonstrated that fossil fields can reside and be stable over long time-scales inside radiative stars, and if observed from the surface would look like the large-scale structures detected a the surface of A/B stars.

\subsection{The BinaMIcS project}

The fossil field hypothesis naively implies that all high-mass stars should display magnetic fields at their surface. This clearly disagrees with observations that find a magnetic incidence fraction of 5 to 10\%. A natural explanation would be the existence of fundamental differences in the initial conditions of star forming regions (e.g. local density, local magnetic  field strength, etc.). An efficient way to test this hypothesis is to study the magnetic properties of a large number ($\sim150$) of close binary systems, containing two stars formed at the same time and from the same environment. This is one of the aim of the recently started BinaMIcS\footnote{http://lesia.obspm.fr/binamics/} project (Binarity and Magnetic Interaction in various classes of Stars). Among other objectives, this project will acquire about 300 high-resolution spectropolarimetric spectra of 150 close binary systems, thanks to large programs obtained on ESPaDOnS (PI: Alecian/Wade) and Narval (Neiner). These data will allow us to determine the incidence and to characterise the magnetic fields of both components of these binaries. As the stars selected in this project are expected to be coeval, BinaMIcS will therefore help us to disentangle initial condition effects from other effects (e.g. early evolutionary, or rotational) on the low magnetic incidence. This project is one further step towards our understanding of the fossil field theory (Alecian, Wade, Mathis, Neiner et al., in prep.).

\subsection{The IMTTS project}

The Herbig Ae/Be survey, introduced above, led to the conclusion that magnetic fields, as observed in Herbig Ae/Be and in Ap/Bp stars, must have been shaped during earlier phases of the stellar evolution. The distribution of the Herbig Ae/Be stars in the HR diagram place all of them in the fully radiative phase or with a small convective core. To study the earliest phases, where convection is active, we need to concentrate on the intermediate-mass T Tauri stars (IMTTS). We have recently started a new project aiming at determining the magnetic properties in a sample of about 50 IMTTS, using ESPaDOnS, HARPSpol and CRIRES observations. This project will allow us to characterise the magnetic fields at the surface of the phase on both sides of the limit in the HR diagram predicting the disappearing of the convective envelope. This project will therefore allow us to test the hypothesis that fossil fields are formed during the convective phases of the star formation, and then relax into radiative interiors (Alecian, Hussain, Morin et al., in prep.). 

\section{A tantalising view of the PMS evolution of a magnetic intermediate mass star}

Our current view of the evolution of the magnetic field in a PMS intermediate mass star is schematically described in Fig. 1. During phase 1 (Fig. 1), the star is fully convective. Main-sequence fully convective M stars show large-scale magnetic fields in their surface, much less complex than in partially convective M stars (Morin et al. 2010). By extension we propose that during phase 1, the magnetic fields, generated in the fully convective interior could have a large-scale structure. Then when a radiative core appear (phase 2), a solar-type dynamo occurs in the convective envelope, and complex fields should be observed on the surface, while the field originally created in phase 1 is relaxing into the radiative core. Once the star reaches phase 3 (the fully radiative phase), dynamo has stopped, and the relaxed large-scale fossil field only is observable on the surface. Finally, we are now wondering in the later stage of the PMS evolution, when the convective core is appearing (phase 4), if an interaction of the dynamo created in the core, and the relaxed fossil field in the radiative envelope could occur. Featherstone et al. (2009) have performed some simulations to study such an interaction and they find such an interaction could occur, resulting in a change in the obliquity of the fossil field. Such a phenomenon seem to have been observed in the Herbig Ae star HD 190073 (Alecian et al. 2013a). However measuring such an effect is very subtle, and additional measurements are under analysis to confirm this detection. However, with or without confirmed observations of this effect, it would well be that when Herbig Ae/Be stars cross the limit between phase 3 and phase 4, the fossil field is slightly affected.

\bigskip
{\it Acknowledgements.} Some of the work described in this contribution have been undertaken in collaboration with many other scientists and especially with C. Catala, G.A. Wade, J. Landstreet, S. Mathis, C. Neiner, G. Hussain, J. Morin, and have been partially founded by PNPS.


\begin{thebibliography}{99}

\bibitem{alecian08a}
{Alecian E., Catala C., Wade G. A., Donati J.-F., Petit P., et al.} 2008a, MNRAS, 385, 391

\bibitem{alecian08b}
{Alecian E., Wade G. A., Catala C., Bagnulo S., B\"ohm T., Bohlender D., Bouret J.-C., Donati J.-F., Folsom C. P., Grunhut J., Landstreet J. D.} 2008, A\&A, 481, 99

\bibitem{alecian09}
{Alecian E., Wade G. A., Catala C., Bagnulo S., Bšhm T., Bouret J.-C., Donati J.-F., Folsom C. P., Grunhut J., Landstreet J. D.} 2009, MNRAS 400, 354

\bibitem{alecian13a}
{Alecian E., Neiner C., Mathis S., Catala C., Kochukhov O., Landstreet J.} 2013, A\&A 549, 8

\bibitem{alecian13b}
{Alecian E., Wade G. A., Catala C., Grunhut J. H., Landstreet J. D., Bagnulo S., B\"ohm T., Folsom C. P., Marsden S., Waite I.} 2013, MNRAS 429, 1001

\bibitem{bailey12}
{Bailey J.D., Grunhut J.,Shultz M., Wade G., Landstreet J. D., et al.} 2012, MNRAS, 423, 328

\bibitem{behrend01}
{Behrend R., Maeder A.} 2001, A\&A 373, 190

\bibitem{borra82}
{Borra E. F., Landstreet J. D., Mestel L.} 1982, ARA\&A, 20, 191

\bibitem{braithwaite04}
{Braithwaite J., Spruit H. C.} 2004, Nature, 431, 819

\bibitem{braithwaite06}
{Braithwaite J., Nordlund \AA} 2006, A\&A 450, 1077

\bibitem{catala93}
{Catala C., B\"ohm T., Donati J.-F., Semel M.} 1993, A\&A 278, 187

\bibitem{catala99}
{Catala C., Donati J. F., B\"ohm T., Landstreet J., Henrichs H. F., et al.} 1999, A\&A 345, 884

\bibitem{catala07}
{Catala C., Alecian E., Donati J.-F., Wade G. A., Landstreet J. D., B\"ohm T., Bouret J.-C., Bagnulo S., Folsom C., Silvester J.} 2007, A\&A, 462, 293

\bibitem{dewinter97}
{de Winter D., Koulis C., Th\'e P. S., van den Ancker M. E., P\'erez M. R., Bibo E. A.} 1997, A\&AS 121, 223

\bibitem{donati97}
{Donati J.-F., Semel M., Carter B. D., Rees D. E., Collier Cameron A.} 1997, MNRAS 291, 658

\bibitem{donati09}
{Donati J.-F., Landstreet J. D.} 2009, ARA\&A, 47, 333

\bibitem{duez10a}
{DuezV., Mathis S.} 2010, 517, 58

\bibitem{duez10b}
{Duez V., Braithwaite J., Mathis S.} 2010, ApJ 724L, 34

\bibitem{featherstone}
{Featherstone N. A., Browning M. K., Brun A.S., Toomre J.} 2009, ApJ 705, 1000

\bibitem{morin10}
{Morin J., Donati J.-F., Petit P., Delfosse X., Forveille T., Jardine M. M.}, 2010, MNRAS 407, 2269

\bibitem{moss01}
{Moss D.} 2001, Magnetic Fields across the Hertzspring-Russel Diagram, ASP Conference Series, 248, 305

\bibitem{noyes84}
{Noyes R. W., Hartmann L. W., Baliunas S. L., Duncan D. K., Vaughan A. H.} 1984, ApJ, 279, 763

\bibitem{palla90}
{Palla F.; Stahler S. W.} 1990, ApJ, 360, 47

\bibitem{park02}
{Park B.-G., Sung, H.} 2002, AJ, 123, 892

\bibitem{petit13}
{Petit V., Owocki S. P., Wade G. A., Cohen D. H., Sundqvist J. O., et al.} 2013, MNRAS, 429, 398

\bibitem{power08}
{Power J., Wade G. A., Auri\`ere M., Silvester J., Hanes D.} 2008, CoSka 38, 443

\bibitem{prendergast56}
{Prendergast K. H.} 1956, ApJ 123, 498

\bibitem{schussler78}
{Sch\"u{\ss}ler M., P\"aler A.}, 1978, A\&A, 68, 57

\bibitem{stahler83}
{Stahler S.W.} 1983, ApJ, 274, 822

\bibitem{sung97}
{Sung H., Bessell M. S., Lee S.-W.} 1997, AJ 114, 2644

\bibitem{the94}
{Th\'e P. S., de Winter D., Perez, M. R.} 199', A\&AS 104, 315

\bibitem{vieira03}
{Vieira S. L. A., Corradi W. J. B., Alencar S. H. P., Mendes L. T. S., Torres C. A. O., Quast G. R., Guimar‹es M. M., da Silva L.} 2003, AJ, 126, 2971

\bibitem{wade05}
{Wade G. A., Drouin D., Bagnulo S., Landstreet J. D., Mason E., Silvester J., Alecian E., Bšhm T., Bouret J.-C., Catala C., Donati J.-F.} 2005, A\&A, 442L, 31

\end{thebibliography}
\end{document}